# Phenomena in bubbles cluster

Ion Simaciu[a,*], Zoltan Borsos[b,**], Viorel Drafta[c] and Gheorghe Dumitrescu[d]

[a] Retired lecturer, Petroleum-Gas University of Ploieşti, Ploieşti 100680, Romania

[b] Petroleum-Gas University of Ploieşti, Ploieşti 100680, Romania

[c] Independent researcher

[d] High School Toma N. Socolescu, Ploieşti, Romania

E-mail: [*] isimaciu@yahoo.com; ** borzolh@upg-ploiesti.ro

**Abstract**

Following some previous papers, we continue in this paper to give proof of the analogy between the acoustic world and the electromagnetic world. Hence we derive the expressions for: the scattering–scattering force (the electro-acoustic force) between the bubbles found in the inner cluster and the bubble found out of the cluster, the scattering–absorption force (the gravito-acoustic forces) between the bubbles found in the inner cluster, the gravito-acoustic forces between the cluster as a whole and an outer bubble, the temperature corresponding to the translational motion of the bubbles in the inner cluster and the average pressure of the acoustic radiation. The results of our calculus led us to find out that the absolute value of the average pressure of the acoustic radiation around a bubble is equal to the density of the energy of the electro-acoustic field, Eq. (10). This is the most important result of this paper. These densities are the densities of the energy corresponding to the oscillation of the bubble (the liquid around the bubble) that are involved in the interaction phenomenon between two oscillating bubbles. We have also demonstrated that the gravito-acoustic forces, at resonance, between the bubbles inside the cluster, generated by the absorption of energy in the bubbles, are proportional to the square of the virtual masses of the bubbles, Eqs. (29) and (30).

## 1. Introduction

During time a lot of papers and experiments were focused on proving the analogy between the systems and phenomena of the acoustic world and of the electromagnetic world. We remind some of phenomena studied: the interaction of two oscillating bubbles and the translational motion [1 - 5, 23, 30 - 34], the dynamics of a system of several bubbles (filament and spherical cluster) and the coupling of oscillations [6 - 8], the acoustic Casimir effect [9 - 12], the interaction between vortices and bubbles [13 - 15], the formation and properties of antibubbles [16, 17], the acoustic lens effect for an oscillating bubble and a cluster [18 - 20], the interaction between droplets, particles solids and bubbles [24], various other interactions involving bubbles [35, 36], the sonoluminescence phenomenon [37], etc. The aim of this paper is to study other phenomena related to a multi-bubble system. In the second part, we complete the study of the electro-acoustic interaction (scattering-scattering interaction) of bubbles found in a cluster. That is, we infer a Gaussian law for this interaction and highlight the hydromechanical significance of the intensity of the electro-acoustic field. In the third section we take over some of the previous approaches to obtain a gravito - acoustic force (scattering-absorption force) between the bubbles in the cluster. In the fourth section we get up our study to the electro-acoustic and gravito - acoustic forces (the scattering-scattering and scattering-absorption forces) between an outer bubble and a cluster. We have reverted to the more general naming of the electro-acoustic forces as scattering-scattering forces because, as we will show in a future paper, these intra-cluster forces are smaller than the scattering-absorption forces (gravito-acoustic forces) at the same distance. What happens when a cluster is compressed and how its temperature evolve are described in the fifth section. There is also an approach of a dumb hole corresponding to the cluster.

## 2. Scattering–scattering force in a cluster: interpretation and significance

### 2.1. Scattering–scattering force and Gauss's law

In the paper Mach's Principle in the Acoustic World [8] we have analyzed the electro-acoustic force (the scattering–scattering force) between two bubbles found in the inner cluster. Then in the paper Interaction of a bubble and a bubble cluster in an acoustic field [7] we found out the shape of the electro-acoustic force between two bubbles, found in the inner cluster, to be

$$\vec{F}_{2B}^{(1)} = \frac{\rho}{4\pi r^3}\left\langle \left(\dot{V}_2\right)^2 \right\rangle \vec{r}, \qquad (1)$$

that is Eq. (10) corrected from the paper [7]. The expressions of this force are given in Eqs. (27 -29) from the paper [8]. If we total this force for all the bubbles found in the inner sphere of radius $r_0$, then the result will be the force with which these bubbles act on a single bubble which is on the surface of the sphere, Eq. (11) from the paper [7]. This expression is analogous to the electrostatic force with which charged particles found in a sphere of radius $r_0$ act on a particle which is on the surface of the sphere, i.e.

$$\vec{F}_{2B}^{(2)} = \frac{N\rho r_0}{4\pi r_{clust}^3}\left\langle V_2 \ddot{V}_2 \right\rangle \vec{e}_{r0} = \frac{-Ne_{eaN}^2 r_0}{r_{clust}^3}\vec{e}_{r0}. \qquad (2)$$

The analogy occurs when $\rho\left\langle V_2\ddot{V}_2\right\rangle/(4\pi) = e_{eaN}^2$, where $e_{eaN}$ will be labelled the electro-acoustic charge of the inner cluster.

Since the electrostatic force is a force invers proportional to the square of the distance, $\vec{F}_e = -q^2\vec{r}/(4\pi\varepsilon_0 r^3)$, then the particles found in the sphere of radius $r_0$, $N_0 = (4\pi r_0^3/3)(3N/4\pi R_c^3) = Nr_0^3/R_c^3$, act on a charged particle on the surface of the sphere, according to Gauss's law [21, Ch. 4] as if it were a charge with a single-value $q_{N_0} = qN_0 = Nqr_0^3/R_c^3$ placed in the center of the sphere, $\vec{F}_{eN_0} = -qq_{N_0}\vec{r}_0/(4\pi\varepsilon_0 r_0^3) = -Nq^2\vec{r}_0/(4\pi\varepsilon_0 R_c^3) = -Ne^2\vec{r}_0/R_c^3$ (with $e^2 = q^2/(4\pi\varepsilon_0)$ the square of the electrostatic charge in vacuum). Except for the sign (the forces between identical electric charges are repulsive) this electrostatic relation is analogous to Eq. (2). Also, according to Gauss's law, the force exerted by charged particlesfound between the sphere of radius $r_0$ and the sphere of radius $R_c$ on a particle found on the sphere of radius $r_0$ or less, $r < r_0$, is zero. The Newtonian gravitational field and the gravito - acoustic field have the same properties. We will use this property to calculate the average of the potential energy of the bubbles in the cluster

## 2.2. Electro-acoustic field intensity, mechanical significance

When we expressed the force exerted between two bubbles in our previous papers, we used only the term proportional to the radial acceleration, $\ddot{R}$, which was a part of the expression of the pressure induced by the oscillation of the bubble [1],

$$p_b(r,t) = \frac{\rho \ddot{V}}{4\pi r} - \frac{\rho \dot{R}^2}{2}\left(\frac{R}{r}\right)^4 = \frac{\rho R^2 \ddot{R}}{r} + \frac{2\rho R \dot{R}^2}{r} - \frac{\rho \dot{R}^2}{2}\left(\frac{R}{r}\right)^4 \cong \frac{\rho R^2 \ddot{R}}{r}. \quad (3)$$

The aim was to keep only the terms in the first order in the dimensionless amplitude of the oscillations $a \leftrightarrow \varepsilon$ [2, 4].

The average induced pressure corresponding to the bubble oscillations is

$$\langle p_b(r,t) \rangle = \frac{\rho \langle R^2 \ddot{R} \rangle}{r} + \frac{2\rho \langle R \dot{R}^2 \rangle}{r} - \frac{\rho \langle \dot{R}^2 R^4 \rangle}{2r^4}. \quad (4)$$

Since

$$R(t) = R_0(1 + x(t)) = R_0\left[1 + a\cos(\omega t + \varphi)\right], \quad (5)$$

then: $\dot{R} = R_0 \dot{x} = -R_0 \omega a \sin(\omega t + \varphi) = -R_0 \omega a \sin\alpha$, $\ddot{R} = R_0 \ddot{x} = -R_0 \omega^2 a \cos\alpha$. Therefore, the expression of the averaged pressure (4) becomes:

$$\langle p_b(r,t) \rangle = \frac{\rho R_0^3}{r}\langle(1+x)^2 \ddot{x}\rangle + \frac{2\rho R_0^3}{r}\langle(1+x)\dot{x}^2\rangle - \frac{\rho R_0^6 \langle(1+x)^4 \dot{x}^2\rangle}{2r^4} =$$

$$\frac{\rho R_0^3}{r}\left[\langle\ddot{x}\rangle + 2\langle x\ddot{x}\rangle + \langle x^2\ddot{x}\rangle + 2\langle\dot{x}^2\rangle + 2\langle x\dot{x}^2\rangle\right] - \quad (6)$$

$$\frac{\rho R_0^6}{2r^4}\left[\langle\dot{x}^2\rangle + 4\langle x\dot{x}^2\rangle + 6\langle x^2\dot{x}^2\rangle + 4\langle x^3\dot{x}^2\rangle + \langle x^4\dot{x}^2\rangle\right].$$

One can use the following equalities: $\langle\ddot{x}\rangle = -a\omega^2\langle\cos\alpha\rangle = 0$, $\langle x\ddot{x}\rangle = -a^2\omega^2\langle\cos^2\alpha\rangle = -a^2\omega^2/2 = -\langle\dot{x}^2\rangle$, $\langle x\dot{x}^2\rangle = a^3\omega^2\langle\cos\alpha\sin^2\alpha\rangle = 0$, $\langle x^2\dot{x}^2\rangle = a^4\omega^2\langle\cos^2\alpha\sin^2\alpha\rangle = a^4\omega^2/8$, $\langle x^2\ddot{x}\rangle = -a^3\omega^2\langle\cos^3\alpha\rangle = 0$, $\langle x^3\dot{x}^2\rangle = a^5\omega^2\langle\cos^3\alpha\sin^2\alpha\rangle = 0$ and $\langle x^4\dot{x}^2\rangle = a^6\omega^2\langle\cos^4\alpha\sin^2\alpha\rangle = a^6\omega^2/16$, to be used into (6), and then follows

$$\langle p_b(r,t) \rangle = p_b(r) = \frac{-\rho R_0^6 a^2 \omega^2}{4r^4}\left(1 + \frac{3a^2}{2} + \frac{a^4}{8}\right) \cong \frac{-\rho R_0^6 a^2 \omega^2}{4r^4} < 0. \quad (7)$$

This pressure is due to the wave scattered by the oscillating bubble, or likewise, due the waves emitted by the oscillating bubble. The meaning of this result is obtained if we consider that the total pressure averaged in time and outside the bubble is the sum of the hydrostatic pressure $p_0$ and the induced pressure averaged in time due to the oscillations of the bubbles

$$p_t = p_0 + p_b(r) = p_0 - \frac{\rho R_0^6 a^2 \omega^2}{4r^4} < p_0, \quad (8)$$

and that is that the time-averaged total pressure around the oscillating bubble is less than the hydrostatic pressure $p_0$.

The total energy corresponding to this time-averaged scattered radiation pressure is

$$\langle E_b \rangle = \int_{R_0}^{R_c} \langle w_b \rangle 4\pi r^2 dr = \int_{R_0}^{\infty} p_b(r) 4\pi r^2 dr = \pi \rho R_0^6 a^2 \omega^2 \int_{R_0}^{\infty} \frac{dr}{r^2} \cong \pi \rho R_0^6 a^2 \omega^2 \left(\frac{1}{R_0}\right) = \pi \rho R_0^5 a^2 \omega^2 = E_{kb}, \quad (9)$$

with $w_b = -p_b(r)$ [21, 22] and this is also the kinetic energy of oscillation, of the liquid around the bubble, averaged over time.

The average density of the energy due to the oscillation is equal to the density of the energy [21, 22] of the electro-acoustic field of a cluster of bubbles, which is written in the CGS Gaussian units system

$$\langle w_{bN} \rangle = w_{eaN} = \frac{E_{eaN}^2}{8\pi} = \frac{e_{eaN}^2}{8\pi r^4} = \frac{|F_{BN}|}{8\pi r^2} = \frac{\rho R_0^6 a_N^2 \omega^2}{4 r^4}, \quad (10)$$

with $e_{eaN}^2$ the square of the acoustic charge for the bubbles in the cluster, according to relation (2) in this paper and the relation (84) of paper [4] or relation (30) of paper [39]:

$$F_{BNss}(r) = \frac{-2\pi\rho\omega^2 R_0^6 a_N^2}{r^2} \cong \frac{-2\pi\omega^2 A^2 R_0^2}{\rho r^2 \left[ (\omega^2 N_c - \omega_0^2)^2 + 4\beta_s^2 \omega^2 \right]} = \frac{-e_{eaN}^2}{r^2},$$

$$e_{eaN}^2 = 2\pi\rho\omega^2 R_0^6 a_N^2 = \frac{2\pi\omega^2 A^2 R_0^2}{\rho \left[ (\omega^2 N_c - \omega_0^2)^2 + 4\beta_s^2 \omega^2 \right]},$$

$$F_{BNssr}(r) = \frac{-2\pi\rho\omega^2 R_0^6 a_{Nr}^2}{r^2} \cong \frac{-2\pi A^2 \rho u^2 R_0^4 N_c^2}{r^2 p_{eff}^2} = \frac{-e_{eaNr}^2}{r^2}, \quad (11)$$

$$e_{eaNr}^2 = \frac{2\pi A^2 \rho u^2 R_0^4 N_c^2}{p_{eff}^2}.$$

The total energy of the electro-acoustic field around a bubble, considering the energy density expression (10), is

$$E_{eaN} = \int_{R_0}^{R_c} w_{eaN} 4\pi r^2 dr \cong \pi\rho R_0^5 a_N^2 \omega^2 = E_{bN} = E_{bkN}. \quad (12)$$

Relation (12) highlights the significance of the electro-acoustic field energy as the average oscillation energy due to the oscillation of the liquid around the oscillating bubble. This energy is interpreted by an outside observer as the rest energy of the oscillating bubble system, like how the energy of the electrostatic field energy of a charged particle without spin is interpreted as the rest energy of the particle.

## 3. Cluster scattering–absorption force

In this section of the paper, we will infer the expression of the gravito-acoustic force (the scattering–absorption force) between two bubbles found inner of a cluster of bubbles. We use Eq. (12), from the paper [5], (i.e. the expression of the gravito-acoustic force), written under the conditions: $A \to A_N$, $\omega_0 \to \omega_N$ and $\beta \to \beta_N$

$$F_{BNsa}(r) = \frac{-16\pi R_0^2 \omega^4 A_N^2 \beta_{sN} \beta_{aN}}{r^2 \rho \left[ (\omega^2 - \omega_N^2)^2 + 4\beta_{sN}^2 \omega^2 \right]^2}. \quad (13)$$

When replacing parameters $A_N = A/(1+N_c)$, $\omega_N = \omega_0/\sqrt{1+N_c}$ and $\beta_N = \beta/(1+N_c)$ given by the expressions (16) from the paper [8], it follows that:

$$F_{BNsa}(r) = \frac{-16\pi R_0^2 \omega^4 A^2 \beta_s \beta_a}{r^2 \rho (1+N_c)^4 \left[ \left( \omega^2 - \omega_0^2/(1+N_c) \right)^2 + 4\left[ \beta_s^2/(1+N_c)^2 \right] \omega^2 \right]^2} =$$
$$\frac{-16\pi R_0^2 \omega^4 A^2 \beta_s \beta_a}{r^2 \rho \left[ \left( \omega^2 (1+N_c) - \omega_0^2 \right)^2 + 4\beta_s^2 \omega^2 \right]^2}. \tag{14}$$

With $\beta_s = \omega^2 R_0/(2u)$, Eq. (14) gets.

$$F_{BNsa}(r) = \frac{-16\pi R_0^2 \omega^4 A^2 \beta_s \beta_a}{r^2 \rho \left[ \left( \omega^2 (1+N_c) - \omega_0^2 \right)^2 + 4\beta_s^2 \omega^2 \right]^2} =$$
$$\frac{-8\pi R_0^3 \omega^6 A^2 \beta_a}{r^2 \rho u \left[ \left( \omega^2 (1+N_c) - \omega_0^2 \right)^2 + R_0^2 \omega^6/u^2 \right]^2}. \tag{15}$$

Then substituting in Eq. (15), the damping coefficient of the oscillations, which is

$$\beta_a = \beta_\upsilon + \beta_{th} = \frac{2\mu}{\rho R_0^2} + \frac{2\mu_{th}(\omega)}{\rho R_0^2}, \tag{16}$$

one finds two types of gravito - acoustic forces: the gravito-acoustic force due to the absorption of the acoustic energy in the liquid

$$F_{BNsa\mu}(r) = \frac{-8\pi R_0^3 \omega^6 A^2 \beta_\upsilon}{r^2 \rho u \left[ \left( \omega^2 (1+N_c) - \omega_0^2 \right)^2 + R_0^2 \omega^6/u^2 \right]^2} = \frac{-16\pi R_0 \omega^6 A^2 \mu}{r^2 \rho^2 u \left[ \left( \omega^2 (1+N_c) - \omega_0^2 \right)^2 + R_0^2 \omega^6/u^2 \right]^2} \tag{17}$$

and the gravito-acoustic force due to the absorption of the acoustic energy in the vapor and/or gas in the bubbles

$$F_{BNsath}(r) = \frac{-8\pi R_0^3 \omega^6 A^2 \beta_{th}(\omega)}{r^2 \rho u \left[ \left( \omega^2 (1+N_c) - \omega_0^2 \right)^2 + R_0^2 \omega^6/u^2 \right]^2} = \frac{-16\pi R_0 \omega^6 A^2 \mu_{th}(\omega)}{r^2 \rho^2 u \left[ \left( \omega^2 (1+N_c) - \omega_0^2 \right)^2 + R_0^2 \omega^6/u^2 \right]^2}. \tag{18}$$

At resonance, $\omega = \omega_N$, the expression of the force (17) becomes

$$F_{BNsa\mu r}(r) = \frac{-16\pi A^2 \mu u^3}{r^2 \rho^2 R_0^3 \omega_N^6} = \frac{-16\pi (1+N_c)^3 A^2 \mu u^3}{r^2 \rho^2 R_0^3 \omega_0^6} =$$
$$\frac{-16\pi (1+N_c)^3 (A^2 \rho u^2)}{p_{eff}^3} \left( \frac{R_0^3 u \mu}{r^2} \right), \tag{19}$$

With the virtual mass of the bubble $m_b = 2\pi R_0^3 \rho/3$ [2], the expression of the force (19) becomes

$$F_{BNsa\mu}(r) = \frac{-24(1+N_c)^3 m_b A^2 u^3 \mu}{r^2 p_{eff}^3} = \frac{-24(1+N_c)^2 m_N A^2 u^3 \mu}{r^2 p_{eff}^3} \tag{20}$$

and therefore, this force is proportional to the virtual mass $m_b$ or to the equivalent virtual mass (due to the coupling of oscillations), according to Eq. (12) from the paper [8], $m_N = (1+N_c) m_b$.

In what it follows we will analyze the gravito-acoustic force due to the absorption of the acoustic energy in the vapor and/ or gas found in the bubbles. An analytical expression of the thermal damping constant $\beta_{th}(\omega) = 2\mu_{th}(\omega)/(\rho R^2)$ is given in the work Acoustic radiation forces: Classical theory and recent advances [24]

$$\beta_{thi} = \frac{3(\gamma-1)\left[X_i(\sinh X_i + \sin X_i) - 2(\cosh X_i - \cos X_i)\right]\omega_{0i}^2}{2X_i\left[X_i(\cosh X_i - \cos X_i) + 3(\gamma-1)(\sinh X_i - \sin X_i)\right]\omega}, \quad (21)$$

with $X = R_0(2\omega/\chi)^{1/2}$, where $\gamma$ is the ratio of specific heats, $\chi = K/(\rho_g c_P) \cong \mu_g/(\rho_g \gamma) = D_g/\gamma$ is the thermal diffusivity of gas/vapour in bubbles, $K$ is the thermal conductivity, $\mu_g$ is the dynamic viscosity of gas/vapour and $D_o = l_o \upsilon_o /3$ is the coefficient of the diffusion (with $l_g$ the mean free path and $\upsilon_g$ is the mean thermal speed in gas/vapour).

In the paper Acoustic force of the gravitational type [25], relation (21) is an approximation:

a) for the case when $X = R_0(2\omega/\chi)^{1/2} \ll 1$ ($\omega \ll \chi/(2R_0^2) = D_g/(2\gamma R_0^2)$)

$$\beta_{th} \cong \frac{(\gamma-1)X^2 \omega_0^2}{60\gamma\omega} = \frac{(\gamma-1)R_0^2 \omega_0^2}{30\gamma\chi} \quad (22)$$

and b) for the case when $X = R_0(2\omega/\chi)^{1/2} \gg 1$ ($\omega \gg \chi/(2R_0^2) = D_g/(2\gamma R_0^2)$)

$$\beta_{th} \cong \frac{3(\gamma-1)\omega_0^2}{2X\omega} = \frac{3(\gamma-1)\omega_0^2 \chi^{1/2}}{2\sqrt{2\omega}\sqrt{\omega}R_0}. \quad (23)$$

Substituting Eqs. (22) and (23) into Eq.(18), it results:

$$F_{BNsath1}(r) = \frac{-4(\gamma-1)\pi R_0^5 \omega_0^2 \omega^6 A^2}{15\gamma\chi r^2 \rho u\left[\left(\omega^2(1+N_c) - \omega_0^2\right)^2 + R_0^2 \omega^6/u^2\right]^2}, \quad \omega \ll \chi/(2R_0^2) \quad (24)$$

and

$$F_{BNsath2}(r) = \frac{-6\sqrt{2}(\gamma-1)\pi R_0^2 \omega^4 \sqrt{\omega} \omega_0^2 \chi^{1/2} A^2}{r^2 \rho u\left[\left(\omega^2(1+N_c) - \omega_0^2\right)^2 + R_0^2 \omega^6/u^2\right]^2}, \quad \omega \gg \chi/(2R_0^2) \quad (25)$$

At resonance, $\omega = \omega_N$, the above forces acquire the following forms:

$$F_{BNsathr}(r) = \frac{-8\pi u^3 A^2 \beta_{th}(\omega_N)}{r^2 R_0 \rho \omega_N^6} = \frac{-8\pi(1+N_c)^3 u^3 A^2 \beta_{th}(\omega_N)}{r^2 R_0 \rho \omega_0^6} =$$
$$\frac{-8\pi(1+N_c)^3 R_0^5 \rho^2 u^3 A^2 \beta_{th}(\omega_N)}{r^2 p_{eff}^3}, \quad (26)$$

$$F_{BNsathr1}(r) = \frac{-4(\gamma-1)\pi R_0 \omega_0^2 A^2 u^3}{15\gamma\chi r^2 \rho \omega_N^6} = \frac{-4\pi(\gamma-1)(1+N_c)^3 R_0 A^2 u^3}{15\gamma\chi r^2 \rho \omega_0^4} =$$
$$\frac{-4\pi(\gamma-1)(1+N_c)^3 R_0^5 \rho u^3}{15\gamma\chi r^2}\left(\frac{A}{p_{eff}}\right)^2, \quad \omega \ll \chi/(2R_0^2). \quad (27)$$

and

$$F_{BNsathr2}(r) = \frac{-6\sqrt{2}(\gamma-1)\pi\omega_0^2 \chi^{1/2} u^3 A^2}{r^2 R_0^2 \rho \omega_N^7 \sqrt{\omega_N}} = \frac{-6\sqrt{2}\pi(\gamma-1)(1+N_c)^{15/4} \chi^{1/2} u^3 A^2}{r^2 R_0^2 \rho \omega_0^5 \sqrt{\omega_0}} =$$

$$\left[-6\sqrt{2}\pi(\gamma-1)(1+N_c)^{15/4}\right]\left(\frac{A}{p_{eff}}\right)^2 \left(\frac{\rho u^2}{p_{eff}}\right)^{3/4} \frac{R_0^{7/2} \rho \chi^{1/2} u^{3/2}}{r^2}, \quad \omega \gg \chi/(2R_0^2) \tag{28}$$

At resonance, the variable $X(\omega_N) \ll 1$ ($X(\omega_N) = R_0(2\omega_N/\chi)^{1/2} = R_0\left(2\gamma\omega_0/(D_g N_c^{1/2})\right)^{1/2}$ $=\left(6\gamma R_0 p_{eff}^{1/2}/(\rho^{1/2} l_g v_g N_c^{1/2})\right)^{1/2} \ll 1$) for any value of the bubble radius $R_0$, because $l_g \gg R_0$, $N_c \gg 1$ și $p_{eff}/(\rho v_g^2) < 1$. Then it follows that, at resonance, only the expression (27) of the force is valid. With the virtual mass of the bubble $m_b = 2\pi R_0^3 \rho/3$, the expression (27) becomes

$$F_{BNsa\mu_{th}r}(r) = \left(\frac{-m_b^2}{r^2}\right)\left[\frac{3(\gamma-1)(1+N_c)^3 u^3 A^2}{5\pi\gamma\chi\rho R_0 p_{eff}^2}\right]. \tag{29}$$

If we substitute in this expression the equivalent virtual mass of the bubble in the cluster, $m_N$, [8], then it results

$$F_{BNsathr}(r) = \left(\frac{-m_N^2}{r^2}\right)\left[\frac{3(\gamma-1)(1+N_c) u^3 A^2}{5\pi\gamma\chi\rho R_0 p_{eff}^2}\right]. \tag{30}$$

Both formats of the force, at resonance, are proportional to the square of the virtual mass, therefore they are analogous to the gravitational force in the electromagnetic world.

In the expression (30) of the gravito-acoustic force we identify the gravitational acoustic constant

$$G_{aNthr} = (1+N_c)\frac{3(\gamma-1)u^3 A^2}{5\pi\gamma\chi\rho R_0 p_{eff}^2} = (1+N_c)G_{0ath}, \tag{31}$$

where the factor $G_{0ath}$ is the acoustic gravitational constant corresponding to the gravito-acoustic interaction of two free bubbles, according to relation (20a) from the paper [5], with $\beta_{a0} = \beta_{th}(\omega_0)$. With this constant, the gravito-acoustic force in the cluster, at resonance, becomes

$$F_{BNsathr}(r) = \frac{-G_{aNthr} m_N^2}{r^2}. \tag{32}$$

According to the paper Acoustic gravitational interaction revised [5], we can introduce the physical quantity, named the acoustic length of the resonant thermal damping $R_{Nth0}$, starting from the ratio between the gravito-acoustic force, Eq.(13), and the electro-acoustic force in the cluster, Eq.(27), from the paper Mach's Principle in the Acoustic World [8]

$$\frac{F_{BNsa}(r)}{F_{BNss}(r)} = \frac{\dfrac{-8\pi R_0^3 \omega^6 A^2 \beta_a}{r^2 \rho u\left[\left(\omega^2(1+N_c)-\omega_0^2\right)^2 + 4\beta_s^2\omega^2\right]^2}}{\dfrac{-2\pi R_0^2 \omega^2 A^2}{r^2 \rho\left[\left(\omega^2(1+N_c)-\omega_0^2\right)^2 + 4\beta_s^2\omega^2\right]}} = \frac{4R_0 \omega^4 \beta_a}{\left[\left(\omega^2(1+N_c)-\omega_0^2\right)^2 + 4\beta_s^2\omega^2\right]u}. \tag{33}$$

At resonance, $\omega = \omega_N$, Eq. (33) becomes

$$\lim_{\omega \to \omega_N}\left(\frac{F_{BNsa}(r)}{F_{BNss}(r)}\right) = \frac{4R_0\omega_N^4 u\beta_a(\omega_N)}{R_0^2\omega_N^6} = \frac{4u\beta_a(\omega_N)}{R_0\omega_N^2} = \frac{4(1+N_c)u\beta_a(\omega_N)}{R_0\omega_0^2} =$$
$$\frac{4(1+N_c)R_0\rho u\beta_a(\omega_N)}{p_{eff}} = \frac{4(1+N_c)\rho u^2}{p_{eff}}\frac{R_0\beta_a(\omega_N)}{u}. \tag{34}$$

If the limit of the ratio is $R_0/R_{Na0}$, then we can define the acoustic length of the resonant absorption damping $R_{Na0}$ as

$$R_{Na0} = \left(\frac{u}{\beta_a(\omega_N)}\right)\left(\frac{p_{eff}}{4(1+N_c)\rho u^2}\right). \tag{35}$$

For thermal absorption, we can define the acoustic length of the resonant thermal damping $R_{Nth0}$, with $\beta_a(\omega_N) = \beta_{Nth0}$

$$R_{Nth0} = \left(\frac{u}{\beta_{Nth0}}\right)\left(\frac{p_{eff}}{4(1+N_c)\rho u^2}\right). \tag{36}$$

According to Eq. (22), the thermal damping coefficient does not depend on the angular frequency and therefore $\beta_{Nth0} = \beta_{th}$. Then substituting the termal damping coefficient into Eq. (36), it follows

$$R_{Nth0} = \left(\frac{u}{\beta_{th}}\right)\left(\frac{p_{eff}}{4(1+N_c)\rho u^2}\right) = \left(\frac{15\gamma\chi u}{(\gamma-1)R_0^2\omega_0^2}\right)\left(\frac{p_{eff}}{2(1+N_c)\rho u^2}\right) =$$
$$\frac{15}{2(1+N_c)(\gamma-1)}\left(\frac{\gamma\chi}{u}\right) = \frac{5\upsilon_g l_g}{2(1+N_c)(\gamma-1)u}. \tag{37}$$

From Eqs. (31) and (37), results an expression of the acoustic gravitational constant, in the cluster, which is a function of $R_{Nth0}$ (the acoustic length of the resonant thermal damping)

$$G_{aNthr} = \frac{2(1+N_c)(\gamma-1)u}{15\gamma\chi}\frac{9u^2A^2}{2\pi\rho R_0 p_{eff}^2} = \left(\frac{R_0^2 u^2}{R_{Nth0}m_b}\right)\left(\frac{3A^2}{p_{eff}^2}\right) = \left(\frac{R_0^2 u^2}{R_{Nth0}m_N}\right)\left(\frac{3(1+N_c)A^2}{p_{eff}^2}\right). \tag{38}$$

This relation is analogous to Eq. (25) from paper [5].

## 4. Scattering-scattering and scattering-absorption forces outside the cluster

To infer the expression of the electro-acoustic forces between a bubble found in the inner of a cluster and a bubble found outside of the cluster (for identical bubbles, $R_{01} = R_{02} = R_0$), we use the expression of the Bjerknes secondary force [2, 4]:

$$F_{B12}(r,\varphi) = -\frac{2\pi\rho\omega^2 R_{01}^3 R_{02}^3}{r^2}a_1 a_2 \cos(\varphi_2 - \varphi_1) \tag{39}$$

and $\cos(\varphi_2 - \varphi_1)$ given by Eq. (3) from paper [5]

$$\cos(\varphi_2 - \varphi_1) = \frac{(\omega^2 - \omega_1^2)(\omega^2 - \omega_2^2) + 4\beta_1\beta_2\omega^2}{\sqrt{(\omega^2 - \omega_1^2)^2 + 4\beta_{s1}^2\omega^2}\sqrt{(\omega^2 - \omega_2^2)^2 + 4\beta_{s2}^2\omega^2}}. \tag{40}$$

In the above expressions, according to the results found in the paper [7], subscript 1 refers to the parameters of the outer bubble and subscript 2 to the parameters of the bubble found in the inner cluster. Considering the results of papers [5, 8], the dimensionless amplitude $a_1$ has the expression

$$a_1 = \frac{A}{\rho R_0^2 \left[ \left( \omega^2 - \omega_0^2 \right)^2 + 4\beta_s^2 \omega^2 \right]^{1/2}} \tag{41}$$

and the dimensionless amplitude $a_2$ has the expression

$$a_2 \to a_N = \frac{A}{\rho R_0^2 \left[ \left( \omega^2 (1 + N_c) - \omega_0^2 \right)^2 + 4\beta_s^2 \omega^2 \right]^{1/2}}, \tag{42}$$

Eq. (40) gets

$$\cos(\varphi_N - \varphi_1) = \frac{\left( \omega^2 - \omega_0^2 \right)\left( \omega^2 - \omega_N^2 \right) + 4\beta \beta_N \omega^2}{\sqrt{\left( \omega^2 - \omega_0^2 \right)^2 + 4\beta_s^2 \omega^2} \sqrt{\left( \omega^2 - \omega_N^2 \right)^2 + 4\beta_{sN}^2 \omega^2}}. \tag{43}$$

If one performs the following substitutions: $\omega_N^2 = \omega_0^2/(1+N_c)$, $\beta_N = \beta/(1+N_c)$ and $\beta_{sN} = \beta_s/(1+N_c)$ according to Eq. (16) from paper [8], into Eq. (43), then one obtains

$$\cos(\varphi_N - \varphi_1) = \frac{\left( \omega^2 - \omega_0^2 \right)\left( \omega^2 (1+N_c) - \omega_0^2 \right) + 4\beta^2 \omega^2}{\sqrt{\left( \omega^2 - \omega_0^2 \right)^2 + 4\beta_s^2 \omega^2} \sqrt{\left( \omega^2 (1+N_c) - \omega_0^2 \right)^2 + 4\beta_s^2 \omega^2}}. \tag{44}$$

Substituting into Eq. (39), Eqs (41), (42) and (44), with $\beta = \beta_s + \beta_a$, results

$$F_{BNe}(r) = \frac{-2\pi \omega^2 R_0^2 A^2 \left[ \left( \omega^2 - \omega_0^2 \right)\left( \omega^2 (1+N_c) - \omega_0^2 \right) + 4(\beta_s + \beta_a)^2 \omega^2 \right]}{r^2 \rho \left[ \left( \omega^2 - \omega_0^2 \right)^2 + 4\beta_s^2 \omega^2 \right] \left[ \left( \omega^2 (1+N_c) - \omega_0^2 \right)^2 + 4\beta_s^2 \omega^2 \right]} =$$

$$\frac{-2\pi \omega^2 R_0^2 A^2 \left[ \left( \omega^2 - \omega_0^2 \right)\left( \omega^2 (1+N_c) - \omega_0^2 \right) + 4\beta_s^2 \omega^2 \right]}{r^2 \rho \left[ \left( \omega^2 - \omega_0^2 \right)^2 + 4\beta_s^2 \omega^2 \right] \left[ \left( \omega^2 (1+N_c) - \omega_0^2 \right)^2 + 4\beta_s^2 \omega^2 \right]} +$$

$$\frac{-16\pi \omega^4 R_0^2 A^2 \beta_s \beta_a}{r^2 \rho \left[ \left( \omega^2 - \omega_0^2 \right)^2 + 4\beta_s^2 \omega^2 \right] \left[ \left( \omega^2 (1+N_c) - \omega_0^2 \right)^2 + 4\beta_s^2 \omega^2 \right]} + \tag{45}$$

$$\frac{-8\pi \omega^4 R_0^2 A^2 \beta_a^2}{r^2 \rho \left[ \left( \omega^2 - \omega_0^2 \right)^2 + 4\beta_s^2 \omega^2 \right] \left[ \left( \omega^2 (1+N_c) - \omega_0^2 \right)^2 + 4\beta_s^2 \omega^2 \right]}.$$

In Eq. (45), the first relation is the expression of the electro - acoustic force

$$F_{BNess}(r) = \frac{-2\pi \omega^2 R_0^2 A^2 \left[ \left( \omega^2 - \omega_0^2 \right)\left( \omega^2 (1+N_c) - \omega_0^2 \right) + 4\beta_s^2 \omega^2 \right]}{r^2 \rho \left[ \left( \omega^2 - \omega_0^2 \right)^2 + 4\beta_s^2 \omega^2 \right] \left[ \left( \omega^2 (1+N_c) - \omega_0^2 \right)^2 + 4\beta_s^2 \omega^2 \right]} \tag{46}$$

and the second relation is the expression of the gravito - acoustic force, between a bubble found in the inner cluster and an identical bubble found outside the cluster,

$$F_{BNesa}(r) = \frac{-16\pi \omega^4 R_0^2 A^2 \beta_s \beta_a}{r^2 \rho \left[ \left( \omega^2 - \omega_0^2 \right)^2 + 4\beta_s^2 \omega^2 \right] \left[ \left( \omega^2 (1+N_c) - \omega_0^2 \right)^2 + 4\beta_s^2 \omega^2 \right]}. \tag{47}$$

One can obtain another form of the Eq. (47) if we use Eq. (16). This new form of Eq. (47) reveals two types of gravito-acoustic forces:

$$F_{BNesa}(r) = \frac{-16\pi\omega^4 R_0^2 A^2 \beta_s (\beta_\upsilon + \beta_{th})}{r^2 \rho \left[(\omega^2 - \omega_0^2)^2 + 4\beta_s^2 \omega^2\right]\left[(\omega^2(1+N_c) - \omega_0^2)^2 + 4\beta_s^2 \omega^2\right]} =$$

$$\frac{-32\pi\omega^4 A^2 \mu \beta_s}{r^2 \rho^2 \left[(\omega^2 - \omega_0^2)^2 + 4\beta_s^2 \omega^2\right]\left[(\omega^2(1+N_c) - \omega_0^2)^2 + 4\beta_s^2 \omega^2\right]} +$$

$$\frac{-16\pi\omega^4 R_0^2 A^2 \beta_s \beta_{th}}{r^2 \rho \left[(\omega^2 - \omega_0^2)^2 + 4\beta_s^2 \omega^2\right]\left[(\omega^2(1+N_c) - \omega_0^2)^2 + 4\beta_s^2 \omega^2\right]} =$$

$$F_{BNesa\mu}(r) + F_{BNesath}(r). \quad (48)$$

with $F_{BNesa\mu}(r)$ the gravito-acoustic force generated by the absorption of the energy in the liquid and the gravito-acoustic force generated by the absorption of the energy in the gas/vapor of the bubble.

If the angular frequency of the exciting wave is the eigenangular frequency/ natural angular frequency of the bubble found outside the cluster, $\omega = \omega_0$, then the forms of the three forces, with $\beta_s = \omega^2 R_0/(2u)$, must be:

$$F_{BNess0}(r) = \frac{-2\pi R_0^2 \omega_0^2 A^2}{r^2 \rho \left[(\omega_0^2(1+N_c) - \omega_0^2)^2 + 4\beta_{s0}^2 \omega_0^2\right]} =$$

$$\frac{-2\pi R_0^2 A^2}{r^2 \rho \omega_0^2 (N_c^2 + R_0^2 \omega_0^2/u^2)} = \frac{-2\pi R_0^4 A^2}{r^2 p_{eff}(N_c^2 + p_{eff}/(\rho u^2))} \cong \quad (49)$$

$$\frac{-2\pi R_0^4 A^2}{r^2 p_{eff} N_c^2};$$

$$F_{BNesa\mu 0}(r) = \frac{-16\pi A^2 u \mu}{r^2 \rho^2 R_0 \omega_0^4 (N_c^2 + R_0^2 \omega_0^2/u^2)} = \frac{-16\pi R_0^3 A^2 u \mu}{r^2 p_{eff}^2 (N_c^2 + p_{eff}^2/(\rho u^2))} \cong$$

$$\frac{-16\pi R_0^3 A^2 u \mu}{r^2 p_{eff}^2 N_c^2}, \quad (50)$$

$$F_{BNesath0}(r) = \frac{-8\pi R_0^3 A^2 u \beta_{th}(\omega_0)}{r^2 \rho (N_c^2 + R_0^2 \omega_0^2/u^2)} \cong \frac{-8\pi R_0^3 A^2 u \beta_{th}(\omega_0)}{r^2 \rho N_c^2}, N_c^2 >> p_{eff}^2/(\rho u^2). \quad (51)$$

If the angular frequency of the exciting wave is the eigenangular frequency of the bubble found inside the cluster, $\omega = \omega_N$, then the three forms of the forces become:

$$F_{BNessN}(r) = \frac{-2\pi \omega_N^2 R_0^2 A^2}{r^2 \rho \left[(\omega_N^2 - \omega_0^2)^2 + R_0^2 \omega_N^6/u^2\right]} = \frac{-2\pi R_0^2 A^2}{r^2 \rho \omega_N^2 \left[(1 - \omega_0^2/\omega_N^2)^2 + R_0^2 \omega_N^2/u^2\right]} =$$

$$\frac{-2\pi(1+N_c)R_0^2 A^2}{r^2 \rho \omega_0^2 \left[N_c^2 + R_0^2 \omega_0^2/(u^2 N_c)\right]} \cong \frac{-2\pi R_0^2 A^2}{r^2 \rho \omega_0^2 N_c} = \quad (52)$$

$$\frac{-2\pi R_0^4 A^2}{r^2 p_{eff} N_c}, \quad N_c^2 >> R_0^2 \omega_0^2/(u^2 N_c), N_c >> 1;$$

$$F_{BNesa\mu N}(r) = \frac{-16\pi A^2 u \mu}{r^2 \rho^2 R_0 \left[(\omega_N^2 - \omega_0^2)^2 + 4\beta_{sN}^2 \omega_N^2\right]} = \frac{-16(1+N_c)^2 \pi A^2 u \mu}{r^2 \rho^2 R_0 \omega_0^4 \left[N_c^2 + R_0^2 \omega_0^2 / \left[u^2(1+N_c)\right]\right]} \cong$$
$$\frac{-16\pi R_0^3 A^2 u \mu}{r^2 p_{eff}^2}, \quad N_c^2 \gg R_0^2 \omega_0^2 / (u^2 N_c);$$
(53)

$$F_{BNesathN}(r) = \frac{-4\pi \omega_N^2 R_0^2 A^2 \beta_{th}}{r^2 \rho \left[(\omega_N^2 - \omega_0^2)^2 + 4\beta_{sN}^2 \omega_N^2\right] \beta_s} = \frac{-4(1+N_c)^2 (\gamma-1) \pi \omega_N^2 R_0^3 A^2}{15\gamma r^2 \chi \rho \omega_0^2 \left[N_c^2 + R_0^2 \omega_0^2 / (u^2(1+N_c))\right] \beta_s} \cong$$
$$\frac{-4(\gamma-1)\pi R_0^5 A^2 u}{15 r^2 \gamma \chi p_{eff}}, \quad N_c^2 \gg R_0^2 \omega_0^2 / (u^2 N_c).$$
(54)

According to relations (53) and (54), the expressions of the gravito-acoustic forces between a bubble found in the inner cluster and an outer bubble no longer depend on the acoustic Mach number.

The forces between the cluster and the outer bubble can be obtained, according to Eq. (7) from paper [3], be amplifying with the number of bubbles, $N$, the expression of force given by Eqs. (46 - 54).

## 5. Cluster compression, thermal effects

### 5.1. Average pressure due to the oscillations of the bubbles

To find the average pressure in the cluster, due to oscillations of the bubbles, we will use the expression of the pressure around a bubble. The pressure around the bubble, due to bubble volume oscillations [1, 2, 8], found in the inner cluster, is

$$p_{bN}(r,t) = \frac{\rho \ddot{V}_N}{4\pi r} - \frac{\rho \dot{R}_N^2}{2}\left(\frac{R_N}{r}\right)^4 = \frac{\rho R_N^2 \ddot{R}_N}{r} + \frac{2\rho R_N \dot{R}_N^2}{r} - \frac{\rho \dot{R}_N^2}{2}\left(\frac{R_N}{r}\right)^4. \quad (55)$$

The total pressure generated by the $N$ bubbles in the cluster is

$$p_{is} = \rho \sum_{j \neq i} \frac{R_{Nj}^2 \ddot{R}_{Nj}}{r_{ij}} + 2\rho \sum_{j \neq i} \frac{R_{Nj} \dot{R}_{Nj}^2}{r_{ij}} - \frac{\rho}{2} \sum_{j \neq i} \frac{\dot{R}_{Nj}^2 R_{Nj}^4}{r_{ij}^4}. \quad (56)$$

Substituting: $R(t)$ given by Eq. (5) (with the index $N$, corresponding to the bubble inside the cluster), $\dot{R}_N$ and $\ddot{R}_N$, the above expression (56) becomes:

$$p_{is} = \rho \sum_{j \neq i} \frac{R_0^3(1+x_N)^2 \ddot{x}_N}{r_{ij}} + 2\rho \sum_{j \neq i} \frac{R_0^3(1+x_N)\dot{x}_N^2}{r_{ij}} - \frac{\rho}{2}\sum_{j \neq i} \frac{R_0^6(1+x_N)^4 \dot{x}_N^2}{r_{ij}^4} =$$
$$\rho R_0^2 (1+x_N)^2 \ddot{x}_N \sum_{j \neq i} \frac{R_0}{r_{ij}} + 2\rho R_0^2 (1+x_N)\dot{x}_N^2 \sum_{j \neq i} \frac{R_0}{r_{ij}} - \frac{\rho R_0^2 (1+x_N)^4 \dot{x}_N^2}{2}\sum_{j \neq i} \frac{R_0^4}{r_{ij}^4}.$$
(57)

The average pressure, $\langle p_{is} \rangle$, generated by the coupling of the oscillations of the bubbles, is

$$\langle p_{is}\rangle = \rho R_0^2 \left\langle (1+x_N)^2 \ddot{x}_N \right\rangle \sum_{j\neq i} \frac{R_0}{r_{ij}} + 2\rho R_0^2 \left\langle (1+x_N)\dot{x}_N^2 \right\rangle \sum_{j\neq i} \frac{R_0}{r_{ij}} - \frac{\rho R_0^2 \left\langle (1+x_N)^4 \dot{x}_N^2 \right\rangle}{2} \sum_{j\neq i} \frac{R_0^4}{r_{ij}^4} =$$

$$\rho R_0^2 \sum_{j\neq i} \frac{R_0}{r_{ij}} \left[ \langle \ddot{x}_N \rangle + 2\langle x_N \ddot{x}_N \rangle + \langle x_N^2 \ddot{x}_N \rangle + 2\langle \dot{x}_N^2 \rangle + 2\langle x_N \dot{x}_N^2 \rangle \right] - \tag{58}$$

$$\frac{\rho R_0^2}{2} \sum_{j\neq i} \frac{R_0^4}{r_{ij}^4} \left[ \langle \dot{x}_N^2 \rangle + 4\langle x_N \dot{x}_N^2 \rangle + 6\langle x_N^2 \dot{x}_N^2 \rangle + 4\langle x_N^3 \dot{x}_N^2 \rangle + \langle x_N^4 \dot{x}_N^2 \rangle \right].$$

For a free bubble we already have an expression for the pressure, Eqs. (6, 7). Using it, one can rewrite (58) as

$$\langle p_{is} \rangle = \frac{-\rho R_0^2 a_N^2 \omega^2}{4} \left( 1 + \frac{3a_N^2}{2} + \frac{a_N^4}{8} \right) \sum_{j\neq i} \frac{R_0^4}{r_{ij}^4} \cong \frac{-\rho R_0^2 a_N^2 \omega^2}{4} \sum_{j\neq i} \frac{R_0^4}{r_{ij}^4}. \tag{59}$$

For a uniform distribution of bubbles, the sum $\sum^N (R_0^4/r_{ij}^4)$ can be approached through integral calculation, according to the procedure used in the paper [8],

$$\sum_{j\neq i} \frac{R_0^4}{r_{ij}^4} \to R_0^4 \int_{R_0}^{R_c} \frac{n_c dV}{r^4} = R_0^4 \int_{R_0}^{R_c} \frac{4\pi n_c dr}{r^2} \cong \frac{4\pi R_0^4 n_c}{R_0} = \frac{3NR_0^3}{R_c^3} \ll 1. \tag{60}$$

The lower limit of the integral is the radius $R_0$ of the bubble because the waves scattered by the bubble depart from the surface of the bubble.

If we use the above result, with, $N = N_c \left( 2R_c/(3R_0) \right)$ [8] then (59) becomes

$$\langle p_{is} \rangle \cong \frac{-3N\rho R_0^5 a_N^2 \omega^2}{4R_c^3} = \frac{-3NR_0 A^2 \omega^2}{4\rho R_c^3 \left[ \left( \omega^2(1+N_c) - \omega_0^2 \right)^2 + R_0^2 \omega^6/u^2 \right]} =$$

$$\frac{-N_c A^2 \omega^2}{2\rho R_c^2 \left[ \left( \omega^2(1+N_c) - \omega_0^2 \right)^2 + R_0^2 \omega^6/u^2 \right]}. \tag{61}$$

At resonance, $\omega = \omega_N$, the pressure (61) becomes

$$\langle p_{isr} \rangle = \frac{-3NA^2 u^2}{4\rho R_0 R_c^3 \omega_N^4} = \frac{-3NN_c^2 R_0^3 A^2 \rho u^2}{4R_c^3 p_{eff}^2} = \frac{-27N^3 R_0^6 A^2 \rho u^2}{16R_c^6 p_{eff}^2}. \tag{62}$$

The meaning of this result may be obtained if we take into account that the average total pressure outside the bubbles, which are found in the inner cluster, is the sum of the hydrostatic pressure $p_0$ and the average pressure due by the oscillations of the bubbles

$$p_{tc} = p_0 + \langle p_{is} \rangle = p_0 - \frac{-N_c A^2 \omega^2}{2\rho R_c^2 \left[ \left( \omega^2(1+N_c) - \omega_0^2 \right)^2 + R_0^2 \omega^6/u^2 \right]} < p_0, \tag{63}$$

that is, the average total pressure in the cluster is less than the hydrostatic pressure $p_0$.

The total energy corresponding to this time - averaged pressure due the scattered radiation is

$$\langle E_{is} \rangle = \langle w_{is} \rangle V_c = -\langle p_{is} \rangle \frac{4\pi R_c^3}{3} = N \left( \pi \rho R_0^5 a_N^2 \omega^2 \right) = N \left\{ \frac{\pi R_0 A^2 \omega^2}{\rho \left[ \left( \omega^2(1+N_c) - \omega_0^2 \right)^2 + R_0^2 \omega^6/u^2 \right]} \right\}. \tag{64}$$

where we considered that the stress–energy tensors of the field and of the electro-acoustic radiation are analogous to those of the field and of the electromagnetic radiation [21 - Sch.31.8, 22 - Ch.8.] $w_{is} = -p_{is}$.

Also, we will further prove that the energy corresponding to the averaged pressure for a bubble found in the inner cluster is precisely the term in the bracket of the Eq. (64). By averaging relation (55) over time, it follows

$$\langle p_{bN}(r,t) \rangle = \frac{\rho \langle R^2 \ddot{R}_N \rangle}{r} + \frac{2\rho \langle R_N \dot{R}_N^2 \rangle}{r} - \frac{\rho \langle \dot{R}_N^2 R_N^4 \rangle}{2r^4} = -\frac{\rho R_0^6 a_N^2 \omega^2}{4r^4}. \tag{65}$$

The density of the energy corresponding to this pressure is $\langle w_{bin} \rangle = -\langle p_{bN} \rangle = \rho R_0^6 a_N^2 \omega^2 / (4r^4)$. If we put this relation in the form $\langle w_{bin} \rangle = \left[ R_0^4 \left( \rho R_0^2 a_N^2 \omega^2 / 4 \right) \right] / r^4 = R_0^4 \left( w_{0kN} \right) / r^4$ in which we highlighted an average density of kinetic energy of oscillation, $w_{0kN} = \rho R_0^2 a_N^2 \omega^2 / 4 = \rho \langle (\dot{R}_N)^2 \rangle / 2$, then the average energy of oscillation of the fluid around the bubble is

$$E_{bin} = \int_{R_0}^{R_c} \langle w_{bin} \rangle 4\pi r^2 dr = \pi \rho R_0^6 a_N^2 \omega^2 \int_{R_0}^{R_c} \frac{dr}{r^2} = \pi \rho R_0^6 a_N^2 \omega^2 \left( \frac{1}{R_0} - \frac{1}{R_c} \right) \cong \pi \rho R_0^5 a_N^2 \omega^2 = E_{bkin}. \tag{66}$$

The average density of the energy due to oscillation is equal to the density of the energy of the electro - acoustic field [21 – Ch. 28], written in the CGS-Gaussian system of units, of a cluster bubble

$$\langle w_{bin} \rangle = w_{eain} = \frac{E_{eain}^2}{8\pi} = \frac{e_{eain}^2}{8\pi r^4} = \frac{|F_{BN}|}{8\pi r^2} = \frac{\rho R_0^6 a_N^2 \omega^2}{4r^4}, \tag{67}$$

with $e_{eain}^2 = e_{eaN}^2$ (the notation in Eqs. (10, 11) the square of the acoustic charge for the bubbles found in the inner cluster), according to relation (11).

Considering the density of the energy, from Eq. (67), the total energy of the electro-acoustic field around a bubble, is

$$E_{eain} = \int_{R_0}^{R_c} w_{eain} 4\pi r^2 dr \cong \pi \rho R_0^5 a_N^2 \omega^2 = E_{bin} = E_{bkin}. \tag{68}$$

The relation (68) highlights the significance of the energy of the electro - acoustic field as the average energy of the oscillation of the liquid around the oscillating bubble. This energy is interpreted by an outside observer as the rest energy of the oscillating bubble system, analogous to how the energy of the electrostatic field of a charged particle without spin is interpreted as the rest energy of the particle.

The energy of the field of the $N$ bubbles found in the cluster is $E_{bNt} = NE_{bin} = N\left( \pi \rho R_0^5 a_N^2 \omega^2 \right)$, i.e. precisely the expression of the averaged total energy of the radiation, $\langle E_{is} \rangle$, Eq. (64). It follows, as expected, according to the model of interaction of bubbles which oscillate, that radiation scattered by bubble is involved (has an effect/affects) onto phenomenon where coupled oscillators interact [23, 26].

## 5.2. Acoustic temperature

The acoustic temperature, $T_a$, is different from the temperature of the substrate/liquid, $T$, i.e. the temperature of the liquid. The latter is close to the boiling temperature. To prove the difference, we use the virial theorem to estimate the translational kinetic energy of a bubble in the cluster.

According to the virial theorem [27, Sch.10], if the bubbles, in translational motion, do not radiate, then the translational kinetic energy of a bubble in the cluster is equal to half the absolute value of the potential energy

$$E_{bk} = \frac{|E_{bp}|}{2} = \frac{|E_{bpN}|}{2N}, \tag{69}$$

of the bubble involved in the acoustic interaction with the other bubbles in the cluster

Let assume a cluster of bubbles with radius $R_c$ containing $N$ uniformly distributed bubbles ($N = (4\pi R_c^3/3)n_c$). The potential energy of the electro-acoustic interaction of the bubbles found in the inner cluster can be expressed as

$$E_{bpN} = \int_0^{R_c} \frac{(-N(r)e_{eain})d(e_{eain}N(r))}{r} = \frac{-(4\pi)^2 n_c^2 e_{eain}^2}{3} \int_0^{R_c} r^4 \, dr =$$
$$\frac{-3e_{eain}^2}{5R_c}\left(\frac{4\pi R_c^3 n_c}{3}\right)^2 = \frac{-3N^2 e_{eain}^2}{5R_c} = NE_{bp}, \; E_{bp} = \frac{-3Ne_{eain}^2}{5R_c}, \tag{70}$$

according to the (Eq. 60.11) from the paper [28, p. 87]).

In order to calculate the potential energy of a bubble in interaction with the other bubbles we will use the expression for the potential energy between any two bubbles separated by the distance $r_{ij}$ then we will it sum for $N$ bubbles, yielding the expression for the potential energy of a bubble in the cluster, which is

$$E_{bp}^* = \sum_{j \neq i} E_{bpij}^* = \sum_{j \neq i} \frac{-e_{eain}^2}{r_{ij}} = \frac{-e_{eain}^2}{R_0} \sum_{j \neq i} \frac{R_0}{r_{ij}} = N_c \left(\frac{-e_{eain}^2}{R_0}\right). \tag{71}$$

For a continuous bubble distribution, as estimated in the paper [8, Eq. (24) arxiv or Eq. (25) BPI], the summation becomes an integral. With the expression $N_c = 3NR_0/(2R_c)$ the potential energy acquires the following form

$$E_{bp}^* = \frac{3NR_0}{2R_c}\left(\frac{-e_{eain}^2}{R_0}\right) = \frac{-3Ne_{eain}^2}{2R_c}. \tag{72}$$

Comparing Eq. (72) with the expression of the potential energy given by relation (70), it follows that they are in the ratio $E_{bp}^*/E_{bp} = 5/2$.

Then substituting the electro - acoustic charge, Eq. (11), in the expression of the total potential energy, Eqs. (70, 72), it follows:

$$E_{bp} = \frac{-3Ne_{eain}^2}{5R_c} = \frac{-6\pi NA^2 R_0^2 \omega^2}{5\rho R_c \left[(\omega^2 N_c - \omega_0^2)^2 + 4\beta_s^2 \omega^2\right]}, \; E_{bp}^* = \frac{-3Ne_{eain}^2}{2R_c} =$$
$$\frac{-3\pi NA^2 R_0^2 \omega^2}{\rho R_c \left[(\omega^2 N_c - \omega_0^2)^2 + 4\beta_s^2 \omega^2\right]}, \; E_{bpr} = \frac{-4\pi N_c^3 R_0^3 \rho u^2 A^2}{5 p_{eff}^2}, \; E_{bpr}^* = \frac{-2\pi N_c^3 R_0^3 \rho u^2 A^2}{p_{eff}^2}. \tag{73}$$

The average translational kinetic energy of a bubble in the cluster is

$$E_{bk} = \frac{3kT_a}{2}. \tag{74}$$

Substituting Eqs. (74) and (73) in Eq. (69), with $N_c = 3NR_0/(2R_c)$ [8], results

$$T_a = \frac{2E_{bk}}{3k} = \frac{|E_{bp}|}{3k} = \frac{Ne_{eain}^2}{5kR_c} = \frac{2\pi N \omega^2 A^2 R_0^2}{5kR_c \rho \left[ \left(3NR_0\omega^2/(2R_c) - \omega_0^2\right)^2 + 4\beta_s^2 \omega^2 \right]} =$$

$$\frac{2N_c e_{eain}^2}{15kR_0} = \frac{4\pi N_c A^2 R_0 \omega^2}{15k\rho \left[ \left(\omega^2 N_c - \omega_0^2\right)^2 + 4\beta_s^2 \omega^2 \right]}, \tag{75}$$

$$T_a^* = \frac{|E_{bp}^*|}{3k} = \frac{\pi N A^2 R_0^2 \omega^2}{\rho R_c k \left[ \left(\omega^2 N_c - \omega_0^2\right)^2 + 4\beta_s^2 \omega^2 \right]} = \frac{2\pi N_c A^2 R_0 \omega^2}{3\rho k \left[ \left(\omega^2 N_c - \omega_0^2\right)^2 + 4\beta_s^2 \omega^2 \right]},$$

and

$$T_{ar} = \frac{2N_c e_{eainr}^2}{15kR_0} = \frac{4\pi N_c^3 A^2 \rho u^2 R_0^3}{15k p_{eff}^2} = \left(\frac{m_b u^2}{k}\right)\left(\frac{2N_c^3 A^2}{5p_{eff}^2}\right) = \left(\frac{m_b u^2}{k}\right)\left(\frac{27N^3 R_0^3 A^2}{20R_c^3 p_{eff}^2}\right), m_b = \frac{2\pi R_0^3 \rho}{3},$$

$$T_{ar}^* = \frac{|E_{bpr}^*|}{3k} = \frac{2N_c e_{eainr}^2}{3kR_0} = \frac{2\pi N_c^3 A^2 \rho u^2 R_0^3}{3k p_{eff}^2} = \left(\frac{m_b u^2}{k}\right)\left(\frac{N_c^3 A^2}{p_{eff}^2}\right) = \left(\frac{m_b u^2}{k}\right)\left(\frac{27N^3 R_0^3 A^2}{8R_c^3 p_{eff}^2}\right). \tag{76}$$

If we consider that a part of the energy is emitted by radiation, then that part of the energy is transformed into a thermal background of acoustic radiation.

At thermal equilibrium, the temperature of this thermal acoustic radiation is equal to the temperature of the bubble "gas" found in the inner cluster. In this case, each bubble in the inner cluster oscillates under the action of a thermal background. The background has not a monochromatic background of mean pressure, $\langle p_{is} \rangle$, generated by the coupling of bubble oscillations, but a Planck distribution. Thermalization of acoustic radiation, generated by the transalational motion of the bubbles, is done by the Doppler effect [29, Sch. 67]. We will study the phenomenon of thermalization of acoustic radiation in the cluster in a further paper.

### 5.3. Dumb Hole corresponding to a collapsed cluster

In the previous paper Acoustic Black Hole Generated by a Cluster of Oscillating Bubbles [20], we studied the conditions under which the liquid around a cluster becomes an acoustic lens. Also in the same paper we emphasized that a collapsing cluster can become a dumb hole.

In the case of a cluster of identical bubbles, the collapse is due to the electro - acoustic forces. Since the cluster collapse, then the "gas" of the bubbles heats up, generating a thermal pressure. The collapse takes place until a balance is reached between the thermal pressure and the pressure of the electro - acoustic forces, $p(T_a) = p_{ea}$.

Therefore, one can express the pressure due to the electro-acoustic forces, in the case of a uniform distribution of the bubbles, $n_c = \left(3N/4\pi R_c^3\right)$. The pressure due to the electro -acoustic force acting between the bubbles found in the inner sphere of radius $r$ and the bubbles found on the spherical shell of radius $r$ and of thickness $dr$ is

$$dp = \frac{\left(4\pi r^3 n_c e_{eain}\right)\left(4\pi r^2 n_c e_{eain} dr\right)}{3r^2} \frac{1}{4\pi r^2} = \left(\frac{4\pi n_c^2 e_{eain}^2}{3}\right) r dr. \tag{77}$$

Integrating Eq. (77) and substituting the bubble density yields

$$p_{eac} = \frac{4\pi n_c^2 e_{eain}^2}{3} \int_0^{R_c} r dr = \frac{4\pi n_c^2 e_{eain}^2 R_c^2}{6} = \frac{3N^2 e_{eain}^2}{8\pi R_c^4}. \tag{78}$$

Also substituting the expression for the absolute value of the total potential energy, Eqs. (70-72), it results

$$p_{eac} = \frac{3N^2 e_{eain}^2}{8\pi R_c^4} = \frac{5|E_{pbN}|}{8\pi R_c^3} = \frac{5w_{pbN}}{6} = \frac{5w_{eac}}{6}, \quad p_{eac}^* = \frac{|E_{pbN}^*|}{4\pi R_c^3} = \frac{1}{3}\frac{3|E_{pbN}|}{4\pi R_c^3} = \frac{w_{pbN}^*}{3} = \frac{w_{eac}^*}{3}. \tag{79}$$

The thermal pressure, $p(T_a)$, is due to the translational energy and is equal to one third of the density of the kinetic energy

$$p(T_a) = \frac{w_k}{3} = \frac{n_{bc} E_{bk}}{3} = \frac{NE_{bk}}{4\pi R_c^3} = \frac{N|E_{bp}|}{8\pi R_c^3} = \frac{|E_{bpN}|}{8\pi R_c^3}, \quad p^*(T_a) = \frac{N|E_{bp}^*|}{8\pi R_c^3} = \frac{3N^2 e_{eain}^2}{16\pi R_c^4} \tag{80}$$

Finally, substituting the expressions of the potential energies, Eqs. (73), into the expression (80), it results:

$$p(T_a) = \frac{|E_{bpN}|}{8\pi R_c^3} = \frac{3N^2 R_0^2 A^2 \omega^2}{20\rho R_c^4 \left[(\omega^2 N_c - \omega_0^2)^2 + 4\beta_s^2 \omega^2\right]},$$

$$p^*(T_a) = \frac{3N^2 e_{eain}^2}{16\pi R_c^4} = \frac{3N^2 R_0^2 A^2 \omega^2}{8\rho R_c^4 \left[(\omega^2 N_c - \omega_0^2)^2 + 4\beta_s^2 \omega^2\right]},$$

$$p_r(T_a) = \frac{|E_{bpNr}|}{8\pi R_c^3} = \frac{3N^2 N_c^2 R_0^4 A^2 \rho u^2}{20 R_c^4 p_{eff}^2} = \frac{27 N^4 R_0^6 A^2 \rho u^2}{80 R_c^6 p_{eff}^2} = \frac{N_c^4 R_0^2 A^2 \rho u^2}{15 R_c^2 p_{eff}^2}. \tag{81}$$

$$p_r^*(T_a) = \frac{3N^2 e_{eainr}^2}{16\pi R_c^4} = \frac{3N^2 N_c^2 R_0^4 A^2 \omega^2 \rho u^2}{8 R_c^4 p_{eff}^2} = \frac{27 N^4 R_0^6 A^2 \rho u^2}{32 R_c^6 p_{eff}^2} = \frac{N_c^4 R_0^2 A^2 \rho u^2}{6 R_c^2 p_{eff}^2}.$$

.

When comparing these pressures, it turns out that the thermal pressure of the translational moving bubbles is less than the pressure of the electro-acoustic forces by the factor of 5. It occurs from the way of how we inferred the potential energy.

When comparing the forms of thermal pressure of the bubble, Eq. (9), with expressions of the average pressure of the cluster due to the internal radiation, Eqs. (61) and (62), it follows, except for a factor of $1/5$, that the thermal pressure (76) is $N$ times higher than this. The factor $1/5$ arises from the different ways in which we calculated the averages of pressure due to the internal radiation and potential energy. But for $N \gg 1$, the difference between the two quantities becomes insignificant. The mean pressure of the internal radiation is the result of the oscillating motion of the bubbles. The translational movement of the bubbles corresponds to a translational thermal radiation with a much higher intensity. The density of the energy of this translational radiation is of the order of the density of the energy due to the translational kinetic energy. This occurs when bubbles and translational radiation are at equilibrium. Therefore, correspondingly, the translational radiation pressure is of the order of the density of the translational kinetic energy. At the level of our knowledge of the phenomena in the cluster, to date, we can state that it is possible for a cluster to collapse and become a dumb hole, that is, the radius of the collapsed cluster to be smaller than the acoustic radius.

Another issue that arises when studying thermal equilibrium is that of relativistic effects, since systems and phenomena in the acoustic world interact and correlate through acoustic waves. If the translational speed of the bubbles approaches the speed of sound in the liquid, then relativistic effects occur. Therefore, in this case the energy and the momentum must be calculated relativistically. The same relativistic limitation appears for the energy due to the oscillation of the bubble when the amplitude of the excitation wave is large. We will address these issues in a further paper.

## 6. Conclusions and discussions

Through the results of this paper, we have deepened the similarity between electromagnetic world and acoustic world. We inferred expressions of the electro-acoustic force between the bubbles found in the inner cluster and of an outer bubble, of the gravito-acoustic forces occurring in the inner cluster and those occurring due to the interaction of the cluster with an outer bubble, of the temperature corresponding to the translational motion of the bubbles found in the inner cluster and of the average pressure due to the acoustic radiation. The most important result of this paper is to give proof that the absolute value of the averaged pressure of the acoustic radiation around a bubble is equal to the density of the energy of the electrostatic field, Eq. (10). These densities belonging to the liquid around the bubble are the densities of the energy that are involved in the interaction phenomenon between two oscillating bubbles. This property supports the hypothesis that the electro - acoustic charge property and the corresponding interaction measure the ability of bubbles to oscillate radially and to absorb and to scatter acoustic waves [23, 26]. By analogy, electrostatic charge measures the ability of charged particles to absorb and scatter electromagnetic radiation and so is characterized by an internal oscillatory motion [38]. It follows that elementary particles must be nonpoints. The model of a non - point particle, but as an oscillating bubble in the electromagnetic vacuum, must be completed with an internal orbital motion (vortex) for the particle to have zero spin. We will address this model, for bubbles in liquid, in a further paper.